\RequirePackage[pagewise,mathlines]{lineno} 
\documentclass[aps,prl,numerical,linenumbers,superscriptaddress,showpacs,floatfix,reprint]{revtex4-1}
\usepackage[dvipdfm]{graphicx}
\usepackage{amsmath,amssymb,amsfonts}
\usepackage{color}
\usepackage[colorlinks=true,linkcolor=blue,plainpages=false,hypertex]{hyperref}

\def\v2{\mbox{$v_2$}}

\newcommand{\mean}[1]{\left\langle #1 \right\rangle}

\begin{document}

\pagewiselinenumbers

%
\title{ Constraints on models for the initial collision geometry \\
in ultra relativistic heavy ion collisions
}
%
%
%
\author{ Roy~A.~Lacey}
\email[E-mail: ]{Roy.Lacey@Stonybrook.edu}
\affiliation{Department of Chemistry, 
Stony Brook University, \\
Stony Brook, NY, 11794-3400, USA}
\affiliation{Physics Department, Bookhaven National Laboratory, \\
Upton, New York 11973-5000, USA}
\author{Rui Wei} 
\affiliation{Department of Chemistry, 
Stony Brook University, \\
Stony Brook, NY, 11794-3400, USA}
\author{ N.~N.~Ajitanand} 
\affiliation{Department of Chemistry, 
Stony Brook University, \\
Stony Brook, NY, 11794-3400, USA}
\author{ J.~M.~Alexander}
\affiliation{Department of Chemistry, 
Stony Brook University, \\
Stony Brook, NY, 11794-3400, USA}
\author{ X.~Gong}
\affiliation{Department of Chemistry, 
Stony Brook University, \\
Stony Brook, NY, 11794-3400, USA}
\author{ J.~Jia}$^2$
\affiliation{Department of Chemistry, 
Stony Brook University, \\
Stony Brook, NY, 11794-3400, USA}
\affiliation{Physics Department, Bookhaven National Laboratory, \\
Upton, New York 11973-5000, USA}
\author{A.~Taranenko}
\affiliation{Department of Chemistry, 
Stony Brook University, \\
Stony Brook, NY, 11794-3400, USA} 

\author{\\ R. Pak}
\affiliation{Physics Department, Bookhaven National Laboratory, \\
Upton, New York 11973-5000, USA}

\author{Horst St\"ocker}
\affiliation{Institut f\"ur Theoretische Physik, Johann Wolfgang Goethe-Universit\"at \\
             D–60438 Frankfurt am Main, Germany} 

\date{\today}


\begin{abstract}

	Monte Carlo simulations are used to compute the centrality dependence of 
the collision zone eccentricities ($\varepsilon_{2,4}$), for both spherical and deformed 
ground state nuclei, for different model scenarios. Sizable model dependent differences 
are observed. They indicate that measurements of the $2^{\text{nd}}$ and $4^{\text{th}}$ 
order Fourier flow coefficients $v_{2,4}$, expressed as the ratio $\frac{v_4}{(v_2)^2}$, 
can provide robust constraints for distinguishing between different theoretical models 
for the initial-state eccentricity. Such constraints could remove one of the largest 
impediments to a more precise determination of the specific viscosity from 
precision $v_{2,4}$ measurements at the Relativistic Heavy Ion Collider (RHIC).

 
\end{abstract}

\pacs{25.75.-q, 25.75.Dw, 25.75.Ld} 

\maketitle


Energetic collisions between heavy ions at the Relativistic Heavy Ion Collider (RHIC), 
produce a strongly interacting quark gluon plasma (QGP). 
In non-central collisions, the hydrodynamic-like expansion of this 
plasma \cite{Gyulassy:2004zy,Huovinen:2001cy,Teaney:2003kp,Romatschke:2007mq,Hama:2007dq,Song:2007ux} 
results in the anisotropic flow of particles in the plane transverse 
to the beam direction \cite{Lacey:2001va,Snellings:2001nf}. At mid-rapidity, the magnitude of this 
momentum anisotropy is characterized by the even order Fourier coefficients;
\begin{equation}
v_{\rm n} = \mean{e^{in(\phi_p - \Phi_{RP})}}, {\text{  }} n=2,4,.., 
\label{eq:1}
\end{equation}  
where $\phi_{p}$ is the azimuthal angle of an emitted particle, 
$\Phi_{RP}$ is the azimuth of the reaction plane and the brackets denote 
averaging over particles and events. The elliptic flow coefficient $v_2$ 
is observed to dominate over the higher order coefficients in Au+Au 
collisions at RHIC ({\em i.e.}\ $v_n \propto (v_2)^\frac{n}{2}$ and 
$v_2 << 1$) \cite{Adams:2003zg,Lacey:2009xx}.

	The magnitudes and trends of $v_{2,4}$ are known to be sensitive to 
the transport properties of the expanding partonic 
matter \cite{Heinz:2002rs,Teaney:2003kp,Lacey:2006pn,
Romatschke:2007mq,Song:2007ux,Drescher:2007cd,Xu:2007jv,Greco:2008fs,Luzum:2008cw,Chaudhuri:2009hj}. 
Consequently, there is considerable current interest in their use for 
quantitative extraction of the specific shear viscosity, {\em i.e.}\ the ratio 
of shear viscosity $\eta$ to entropy density $s$ of the plasma. 
Such extractions are currently being pursued via comparisons 
to viscous relativistic hydrodynamic simulations \cite{Luzum:2008cw,Song:2008hj,Chaudhuri:2009hj}, 
transport model calculations \cite{Xu:2007jv,Greco:2008fs} and   
hybrid approaches which involve the parametrization of scaling 
violations to ideal hydrodynamic behavior \cite{Drescher:2007cd,Lacey:2006pn,Lacey:2009xx}. 
In all cases, accurate knowledge of the initial eccentricity $\varepsilon_{2,4}$ of 
the collision zone, is a crucial unresolved prerequisite for quantitative 
extraction of $\frac{\eta}{s}$. 

	To date, no direct experimental measurements of $\varepsilon_{2,4}$
have been reported. Thus, the necessary theoretical estimates have 
been obtained by way of the overlap geometry corresponding 
to the impact parameter $b$ of the collision, or the number of 
participants $N_{\text{part}}$ in the collision zone. A robust constraint for 
$N_{\text{part}}$ values can be obtained via measurements of the final hadron multiplicity or 
transverse energy. However, for a given value of $N_{\text{part}}$, the theoretical 
models used to estimate $\varepsilon_{2}$ give results which differ by as much 
as {$\sim 25$\%} \cite{Hirano:2005xf,Drescher:2006pi} -- a difference which leads to 
an approximate factor of two uncertainty in the extracted  
$\eta/s$ value \cite{Luzum:2008cw}. Therefore, an experimental constraint 
which facilitates a clear choice between the different theoretical models is  
essential for further progress toward precise extraction of $\eta/s$. 

	In ideal fluid dynamics, anisotropic flow is directly proportional to the initial 
eccentricity of the collision zone. A constant ratio for the flow coefficients 
$\frac{v_4}{(v_2)^2}\approx 0.5$ is also predicted  \cite{Gombeaud:2009ye}.
It is well established that initial 
eccentricity fluctuations also influence the magnitude 
of $v_{2,4}$ significantly \cite{Alver:2006wh,Hama:2007dq,Hirano:2009ah,Lacey:2009xx,Gombeaud:2009ye},  
{\em i.e.}\ the presence of these fluctuations serve to increase the value  
of $v_{2,4}$. Therefore, one avenue to search for new experimental 
constraints, is to use $\varepsilon_{2,4}$ as a proxy for $v_{2,4}$ and study the model 
dependencies of their magnitudes and trends vs. $N_{\text{part}}$. 
	
 	In this communication we present calculated results of $\varepsilon_{2,4}$ for collisions 
 of near-spherical and deformed isotopes, for the Glauber \cite{Alver:2006wh,Miller:2007ri} 
and the factorized Kharzeev-Levin-Nardi (fKLN) \cite{Lappi:2006xc,Drescher:2007ax} models, 
{\em i.e.}\ the two primary models currently employed for eccentricity estimates. We find 
sizable differences, both in magnitude and trend, for the 
the ratios $\frac{\varepsilon_4}{(\varepsilon_2)^2}$
obtained from both models. This suggests that systematic comparisons of the measurements 
for the $N_{\text{part}}$ dependence of the ratio $\frac{v_4}{(v_2)^2}$  
for these isotopic systems, can give direct experimental constraints 
for these models.  

	Monte Carlo (MC) simulations were used to calculate event averaged 
eccentricities (denoted here as $\varepsilon_{2,4}$) within the framework of 
the Glauber (MC-Glauber) and fKLN (MC-KLN) models, for near-spherical 
and deformed nuclei 
which belong  to an isobaric or isotopic series. 
Here, the essential point is that, for such series, 
a broad range of ground state deformations have been observed for relatively small changes 
in the the number of protons or neutrons \cite{Raman:1987yv,Moller:1993ed}.
For each event, the spatial distribution of nucleons in the colliding nuclei were 
generated according to the deformed Woods-Saxon function:
\begin{equation}
\rho(\mathbf{r})=\frac{\rho_{0}}{1+e^{(\text{r}-R_{0}(1+\beta_{2}Y_{20}(\theta) + \beta_{4}Y_{40}(\theta)))/d}},
\label{Eq2}
\end{equation}
where $R_{0}$ and $d$ are the radius and diffuseness parameters and $\beta_{2,4}$ are 
the deformation parameters which characterizes the density distribution of the 
nucleus about its polarization axis ($z'$). 
%
\begin{figure}[tb]
\includegraphics[width=1.0\linewidth]{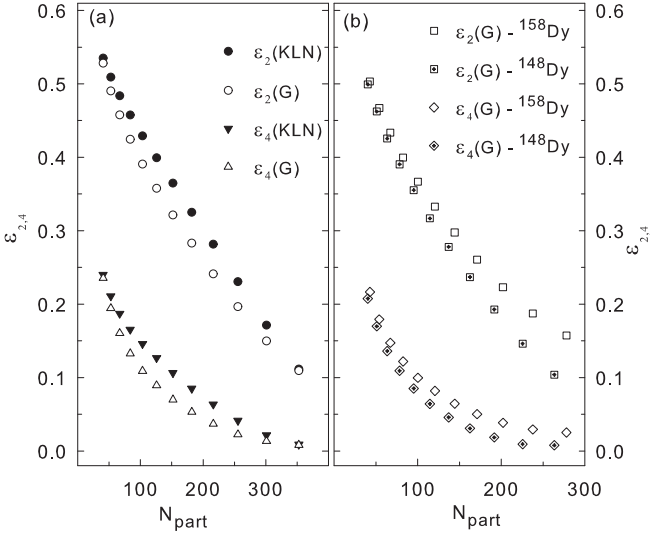}
%
\caption{ Calculated values of $\varepsilon_{2,4}$ vs. $N_{\rm part}$ for 
MC-Glauber (open symbols) and MC-KLN (closed symbols) for Au+Au collisions
(a) and near-spherical $^{148}$Dy and deformed $^{158}$Dy as indicated 
in (b).
}
\label{Fig1}
\end{figure}

	To generate collisions for a given centrality selection, the orientation of the polarization 
axis for each nucleus ($\theta_1, \phi_1$ and $\theta_2, \phi_2$ respectively) was randomly 
chosen in the coordinate frame whose $z$ axis is the beam direction. For each collision,  
the values for $N_{\rm part}$ and the number of binary collisions $N_{\text{coll}}$ were 
determined within the Glauber ansatz \cite{Miller:2007ri}. 
The associated $\varepsilon_{2,4}$ values were then evaluated from the 
two-dimensional profile of the density of sources in the transverse  
plane $\rho_s(\mathbf{r_{\perp}})$, using modified versions of MC-Glauber \cite{Miller:2007ri} and 
MC-KLN \cite{Drescher:2007ax} respectively.

	For each event, we compute an event shape vector $S_{n}$ and the azimuth of 
the the rotation angle $\Psi_n^*$ for $n$-th harmonic of the shape profile \cite{Broniowski:2007ft}; 
  \begin{eqnarray}  
    S_{nx} & \equiv & S_n \cos{(n\Psi^*_n)} = 
    \int d\mathbf{r_{\perp}} \rho_s(\mathbf{r_{\perp}}) \omega(\mathbf{r_{\perp}}) \cos(n\phi), \label{eq:S_x} \\
    S_{ny} & \equiv & S_n \sin{(n\Psi^*_n)} = 
    \int d\mathbf{r_{\perp}} \rho_s(\mathbf{r_{\perp}}) \omega(\mathbf{r_{\perp}}) \sin(n\phi), \label{eq:S_y} \\
    \Psi^*_n & = & \frac{1}{n} \tan^{-1}\left(\frac{S_{ny}}{S_{nx}}\right), \label{eq:S_n-plane}
  \end{eqnarray}
where $\phi$ is the azimuthal angle of each source and the 
weight $\omega(\mathbf{r_{\perp}}) = \mathbf{r_{\perp}}^2$.
The eccentricities were calculated as:
\begin{eqnarray}
\varepsilon_2 = \left\langle \cos 2(\phi - \Psi^*_2) \right\rangle \,\,\,
\varepsilon_4 = \left\langle \cos 4(\phi - \Psi^*_2) \right\rangle
\label{e2e4}
\end{eqnarray}
where the brackets denote averaging over sources, as well as events belonging to 
a particular centrality or impact parameter range. For the MC-Glauber 
calculations, an additional entropy density weight was applied reflecting the 
combination of spatial coordinates of participating nucleons and 
binary collisions  \cite{Hirano:2005xf,Hirano:2009ah} ;
\begin{eqnarray}
\rho_s(\mathbf{r_{\perp}}) \propto \left[ \frac{(1-\alpha)}{2}\frac{dN_{\text{part}}}{d^2\mathbf{r_{\perp}}} + 
                     \alpha \frac{dN_{\text{coll}}}{d^2\mathbf{r_{\perp}}} \right], 
\label{Eq5}
\end{eqnarray}
where $\alpha = 0.14$ was constrained by multiplicity measurements as a 
function of $N_{\text{part}}$ for Au+Au collisions \cite{Back:2004dy}.

		The procedures outlined above (cf. Eqs.~\ref{Eq2} - \ref{Eq5}) ensure that, 
in addition to the fluctuations which stem from the orientation of the 
initial ``almond-shaped'' collision zone [relative to the impact parameter], 
the shape-induced fluctuations due to nuclear deformation are also taken into account. 	
Note that $\varepsilon_{2,4}$ (cf. Eq. \ref{e2e4}) correspond 
to $v_{2,4}$ measurements in the so-called participant 
plane \cite{Alver:2006wh,Miller:2007ri}. That is, the higher harmonic $\varepsilon_{4}$ 
is evaluated relative to the principal axis determined by maximizing the quadrupole moment. 
This is analogous to the measurement of $v_4$ with respect to the $2^{\text{nd}}$ order event-plane 
in actual experiments. One consequence is that the density profile is suppressed, as well as 
the moment for the higher harmonic.

Calculations were performed for a variety of isotopes and isobars with a broad 
range of known $\beta_{2,4}$ values. Here, we show and discuss only a representative 
set of results for $^{197}$Au ($R = 6.38\,\text{fm},\,\beta_2 = -0.13,\,\beta_4 = -0.03$), 
$^{148}$Dy ($R = 5.80\,\text{fm},\,\beta_2 = 0.00,\, \beta_4 = 0.00$) and 
$^{158}$Dy ($R = 5.93\,\text{fm},\, \beta_2 = 0.26,\, \beta_4 = 0.06$) \cite{Raman:1987yv,Moller:1993ed}. 
For these calculations we used the value $d=0.53$ fm. 

	Figure \ref{Fig1}(a) shows a comparison of $\varepsilon_{2,4}$ vs. 
$N_{\text{part}}$ for MC-Glauber (open symbols) and 
MC-KLN (filled symbols) for Au+Au collisions. The filled symbols indicate 
larger $\varepsilon_{2,4}$ values for MC-KLN over most of the 
considered $N_{\text{part}}$ range. The effect of shape deformation is 
illustrated in Fig. \ref{Fig1}(b) where a comparison of $\varepsilon_{2,4}$ vs. 
$N_{\text{part}}$ [for MC-Glauber] is shown for the two Dy isotopes indicated.
Both $\varepsilon_2$ and $\varepsilon_4$ show a sizable increase for the 
isotope with the largest ground state deformation ($^{158}$Dy). This reflects the 
important influence of shape-driven eccentricity fluctuations in collisions of 
deformed nuclei \cite{Shuryak:1999by,Li:1999bea,Heinz:2004ir,Filip:2009zz}. The magnitudes 
and trends of all of these eccentricities are expected to influence the measured values 
of $v_{2,4}$ for these systems. 
%
\begin{figure}[t]
\includegraphics[width=1.0\linewidth]{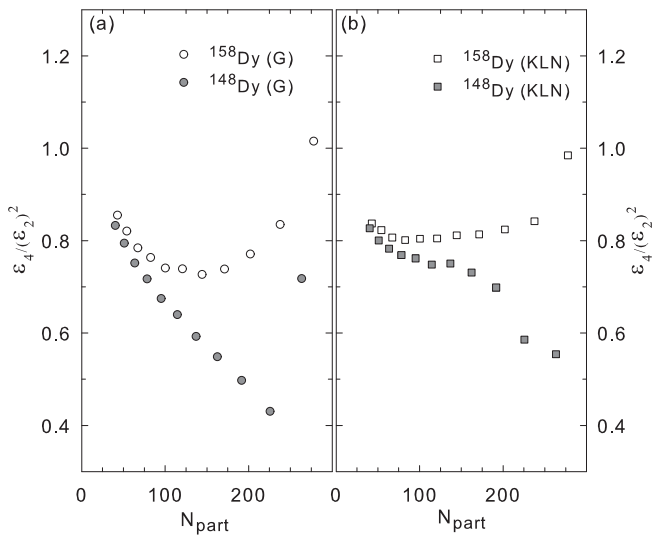}
\caption{Comparison of $\frac{\varepsilon_{4}}{(\varepsilon_{2})^2}$ vs. $N_{\text{part}}$ 
for near-spherical $^{148}$Dy (filled symbols) and deformed $^{158}$Dy (open symbols) 
collisions. 
Results are shown for MC-Glauber (a) and MC-KLN (b) respectively. 
}
\label{Fig2}
\end{figure}

   A priori, the model-driven and shape-driven eccentricity differences shown in Fig. \ref{Fig1}, 
need not be the same for $\varepsilon_{2}$ and $\varepsilon_{4}$. Therefore,
we present the ratio $\frac{\varepsilon_{4}}{(\varepsilon_{2})^2}$ vs. $N_{\text{part}}$, 
for both models in Fig. \ref{Fig2}. The ratios obtained for $^{148}$Dy (near-spherical) and $^{158}$Dy 
(deformed) with MC-Glauber are compared in Fig. \ref{Fig2}(a); the same comparison is given 
in Fig. \ref{Fig2}(b) but for MC-KLN calculations. 
Fig. \ref{Fig2}(a) indicates a significant difference 
between the ratio $\frac{\varepsilon_{4}}{(\varepsilon_{2})^2}$ for $^{148}$Dy
and $^{158}$Dy over the full range of $N_{\text{part}}$ considered. 
This difference stems from additional shape-driven fluctuations present in 
in collisions of $^{158}$Dy, but absent in collisions of $^{148}$Dy. 
The same comparison for MC-KLN results, shown in Fig. \ref{Fig2}(b), 
points to a smaller difference for these ratios, as well as a 
different $N_{\text{part}}$ dependence.  
We attribute this to the difference in the transverse density distributions 
employed in MC-Glauber and MC-KLN.

	For a given value of $N_{\text{part}}$, the measured ratio 
 of the flow coefficients $\frac{v_{4}}{(v_{2})^2}$ for $^{158}$Dy+$^{158}$Dy and 
$^{148}$Dy+$^{148}$Dy collisions, are expected to reflect the magnitude and trend of the ratio 
$\frac{\varepsilon_{4}}{(\varepsilon_{2})^2}$ (note that a constant ratio $\approx 0.5$ 
is predicted for ideal hydrodynamics without the 
influence of fluctuations \cite{Gombeaud:2009ye}). Fig. \ref{Fig2} suggests that a relatively 
clear distinction between fKLN-like and Glauber-like initial collision geometries 
could be made via systematic studies of $\frac{v_4}{(v_2)^2}$ for near-spherical 
and deformed isotopes/isobars. Specifically, a relatively smaller (larger) difference 
between the ratios $\frac{v_4}{(v_2)^2}$ for each isotope, would be expected for fKLN (Glauber) 
initial geometries. Similarly the scaling of $v_{2,4}$ data from 
the isotopic or isobaric pair would be expected only for MC-Glauber or MC-KLN eccentricities. 
Note that the influence of a finite viscosity is expected to be the same for both systems and 
therefore would not change these conclusions.
%
\begin{figure}[t]
\includegraphics[width=1.0\linewidth]{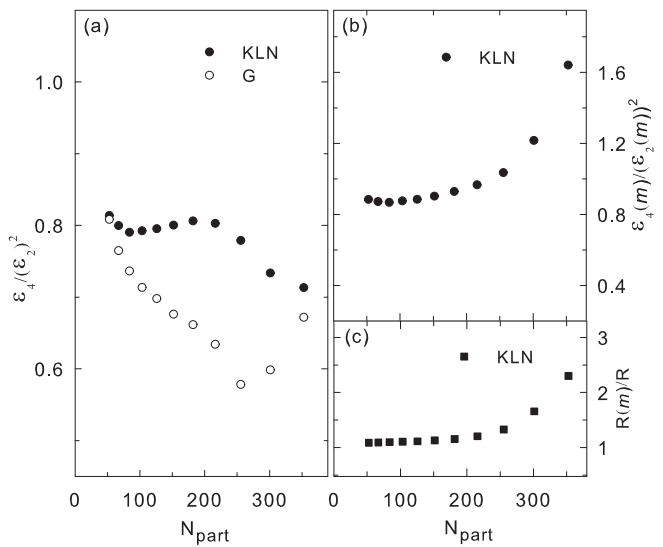} 
\caption{$N_{\text{part}}$ dependence of $\frac{\varepsilon_{4}}{(\varepsilon_{2})^2}$ (a), 
$\frac{\varepsilon_{4}(m)}{(\varepsilon_{2}(m))^2}$ (b) and 
$\frac{\text{R}(m)}{(\text{R})}$ (c) for 
 Au+Au collisions (see text). The open and closed symbols indicate the results from 
MC-Glauber and MC-KLN respectively. 
}
\label{Fig3}
\end{figure}

	The filled symbols in Figs. \ref{Fig2} (a) and (b) also 
suggest a substantial difference in the $\frac{\varepsilon_{4}}{(\varepsilon_{2})^2}$ ratios
predicted by MC-Glauber and MC-KLN respectively, for collisions between 
near-spherical nuclei. This difference is also apparent in
Fig. \ref{Fig3}(a) where the calculated ratios for 
Au+Au ($\beta_2 = -0.13,\,\beta_4 = -0.03$) collisions are shown.
The MC-KLN results (filled circles) indicate a relatively flat 
dependence for $40 \alt N_{\text{part}} \alt 200$, which contrasts with the 
characteristic decrease, for the same $N_{\text{part}}$ range, seen 
in the MC-Glauber results.

	As discussed earlier, each of these trends is expected to influence
the measured ratios of the flow coefficients $\frac{v_{4}}{(v_{2})^2}$. 
Therefore, an experimental observation 
of a relatively flat $N_{\text{part}}$ dependence for $\frac{v_{4}}{(v_{2})^2}$ 
[over the range $40 \alt N_{\text{part}} \alt 200$], could be an indication 
for fKLN-like collision geometries in Au+Au collisions. Such a trend has been
observed in the preliminary and final data sets reported in 
Refs. \cite{Lacey:2009xx,Gombeaud:2009ye,Adare:2010ux} and is consistent 
with the conclusions reached in Ref. \cite{Lacey:2009xx,Heinz:2009cv} that 
the $N_{\text{part}}$ and impact parameter dependence of the eccentricity scaled flow 
coefficients $\frac{v_2}{\varepsilon_2}$ and  $\frac{v_4}{\varepsilon_4}$
favor fKLN-like initial collision geometries.

	 The closed symbols in Figs. \ref{Fig2}(b) and \ref{Fig3}(a) indicate a decreasing 
trend for $\frac{\varepsilon_{4}}{(\varepsilon_{2})^2}$ for near-spherical 
nuclei for $N_{\text{part}} \agt 200$. This decrease can be attributed 
to the fact that, in each event, $\varepsilon_4$ is computed in the 
reference frame which maximizes the quadrupole shape distribution,  
{\em i.e.}\ the so-called {\it participant} frame. In this frame, 
$\varepsilon_4$ can take on positive or negative 
event-by-event values. Consequently, smaller mean 
values are obtained, especially in the most central collisions. 
Fig. \ref{Fig2} shows that the relatively large ground state deformation 
for $^{158}$Dy (open symbols) leads to an increase of 
$\frac{\varepsilon_{4}}{(\varepsilon_{2})^2}$ 
[relative to that for the spherical $^{148}$Dy isotope] which is especially 
pronounced in the most central collisions.
However, Fig. \ref{Fig3}(a) shows that the modest deformation for 
the Au nuclei does not lead to a similarly increasing trend 
for $N_{\text{{part}}} \agt 200$ as implied by 
data \cite{Gombeaud:2009ye,Adare:2010ux}.

	 The relatively flat $N_{\text{part}}$ dependence 
for $\frac{v_{4}}{(v_{2})^2}$, over the range $40 \alt N_{\text{part}} \alt 200$
in Fig. \ref{Fig3}(a), suggests fKLN-like collision geometries. Consequently, 
it is interesting to investigate whether or not the magnitude of the ratios 
for $N_{\text{part}} \agt 200$, can be influenced without 
significant impact on the values for $N_{\text{part}} \alt 200$.
	Figure \ref{Fig3}(b) shows that a large increase of 
$\frac{\varepsilon_{4}}{(\varepsilon_{2})^2}$ can indeed be obtained 
for $N_{\text{part}} \agt 200$ with relatively little change 
in the magnitude and trend of the ratios for $N_{\text{part}} \alt 200$.
This was achieved by introducing a correlation or mixing ($m$) between the 
principal axes of the quadrupole ($\Psi^*_2$) and hexadecapole ($\Psi^*_4$) density profiles 
associated with  $\varepsilon_{2}$ and $\varepsilon_{4}$ respectively. That is,
the orientation of  $\Psi^*_2$ was modified to obtain the new value
$\Psi^{**}_2 = (1-\gamma)\Psi^*_2 + \gamma\Psi^*_4 $, where $\gamma = 0.2$. 
This procedure is motivated by the finding that, in addition to the $v_4$ contributions which 
stem from the initial hexadecapole density profile, experimental measurements 
could also have a contribution from $v_2$ [with magnitude $\propto (v_2)^2$]
\cite{Kolb:2004gi,Broniowski:2007ft}. The correlation
has little, if any, influence on the $\varepsilon_2$ values, but does have 
a strong influence on $\frac{\varepsilon_4}{(\varepsilon_2)^2}$ in the most central 
collisions. This is demonstrated in Fig. \ref{Fig3}(c) where the double ratio 
$\frac{\text{R}(m)}{\text{R}}$ ($\text{R}(m) = \frac{\varepsilon_4(m)}{(\varepsilon_2(m))^2}$ and 
$\text{R} = \frac{\varepsilon_4}{(\varepsilon_2)^2}$) is shown.


In summary, we have presented results for the initial eccentricities $\varepsilon_{2,4}$ for 
collisions of near-spherical and deformed nuclei, for the two primary models currently employed 
for eccentricity estimates at RHIC. The calculated ratios for $\frac{\varepsilon_4}{(\varepsilon_2)^2}$, 
which are expected to influence the measured values of $\frac{v_4}{(v_2)^2}$,  
indicate sizable model dependent differences [both in magnitude and trend] which can be exploited 
to differentiate between the models.
The $\frac{\varepsilon_4}{(\varepsilon_2)^2}$ ratios obtained as a function of $N_{\text{part}}$ 
for Au+Au collisions with the fKLN model ansatz, show trends which are strongly suggestive of the 
measured ratios for $\frac{v_4}{(v_2)^2}$ observed in Au+Au collisions for 
$40 \alt N_{\text{part}} \alt 200$. For more central collisions ($N_{\text{part}} \agt 200$), 
the observed trend is strongly influenced by initial eccentricity fluctuations if a correlation 
between the principal axes of the quadrupole and hexadecapole density profiles
is assumed. New measurements of $\frac{v_4}{(v_2)^2}$ for collisions of near-spherical and deformed 
isotopes (or isobars) are required to exploit these tests. 

{\bf Acknowledgments}
We thank Paul Mantica (MSU/NSCL) for crucial insights on nuclear deformation.
This research is supported by the US DOE under contract DE-FG02-87ER40331.A008 and 
by the NSF under award number PHY-0701487.
 


%
\bibliography{ecc_fluc_x0} 

\begin{thebibliography}{37}%
\makeatletter
\providecommand \@ifxundefined [1]{%
 \@ifx{#1\undefined}
}%
\providecommand \@ifnum [1]{%
 \ifnum #1\expandafter \@firstoftwo
 \else \expandafter \@secondoftwo
 \fi
}%
\providecommand \@ifx [1]{%
 \ifx #1\expandafter \@firstoftwo
 \else \expandafter \@secondoftwo
 \fi
}%
\providecommand \natexlab [1]{#1}%
\providecommand \enquote  [1]{``#1''}%
\providecommand \bibnamefont  [1]{#1}%
\providecommand \bibfnamefont [1]{#1}%
\providecommand \citenamefont [1]{#1}%
\providecommand \href@noop [0]{\@secondoftwo}%
\providecommand \href [0]{\begingroup \@sanitize@url \@href}%
\providecommand \@href[1]{\@@startlink{#1}\@@href}%
\providecommand \@@href[1]{\endgroup#1\@@endlink}%
\providecommand \@sanitize@url [0]{\catcode `\\12\catcode `\$12\catcode
  `\&12\catcode `\#12\catcode `\^12\catcode `\_12\catcode `\%12\relax}%
\providecommand \@@startlink[1]{}%
\providecommand \@@endlink[0]{}%
\providecommand \url  [0]{\begingroup\@sanitize@url \@url }%
\providecommand \@url [1]{\endgroup\@href {#1}{\urlprefix }}%
\providecommand \urlprefix  [0]{URL }%
\providecommand \Eprint [0]{\href }%
\@ifxundefined \urlstyle {%
  \providecommand \doi  [0]{\begingroup \@sanitize@url \@doi}%
  \providecommand \@doi [1]{\endgroup \@@startlink {\doibase
  #1}doi:\discretionary {}{}{}#1\@@endlink }%
}{%
  \providecommand \doi  [0]{doi:\discretionary{}{}{}\begingroup
  \urlstyle{rm}\Url }%
}%
\providecommand \doibase [0]{http://dx.doi.org/}%
\providecommand \Doi [0]{\begingroup \@sanitize@url \@Doi }%
\providecommand \@Doi  [1]{\endgroup\@@startlink{\doibase#1}\@@Doi}%
\providecommand \@@Doi [1]{#1\@@endlink}%
\providecommand \selectlanguage [0]{\@gobble}%
\providecommand \bibinfo  [0]{\@secondoftwo}%
\providecommand \bibfield  [0]{\@secondoftwo}%
\providecommand \translation [1]{[#1]}%
\providecommand \BibitemOpen [0]{}%
\providecommand \bibitemStop [0]{}%
\providecommand \bibitemNoStop [0]{.\EOS\space}%
\providecommand \EOS [0]{\spacefactor3000\relax}%
\providecommand \BibitemShut  [1]{\csname bibitem#1\endcsname}%
\bibitem [{\citenamefont {Gyulassy}\ and\ \citenamefont
  {McLerran}(2005)}]{Gyulassy:2004zy}%
  \BibitemOpen
  \bibfield  {author} {\bibinfo {author} {\bibfnamefont {M.}~\bibnamefont
  {Gyulassy}}\ and\ \bibinfo {author} {\bibfnamefont {L.}~\bibnamefont
  {McLerran}},\ }\Doi {10.1016/j.nuclphysa.2004.10.034} {\bibfield  {journal}
  {\bibinfo  {journal} {Nucl. Phys.},\ }\textbf {\bibinfo {volume} {A750}},\
  \bibinfo {pages} {30} (\bibinfo {year} {2005})}\BibitemShut {NoStop}%
\bibitem [{\citenamefont {Huovinen}\ \emph {et~al.}(2001)\citenamefont
  {Huovinen}, \citenamefont {Kolb}, \citenamefont {Heinz}, \citenamefont
  {Ruuskanen},\ and\ \citenamefont {Voloshin}}]{Huovinen:2001cy}%
  \BibitemOpen
  \bibfield  {author} {\bibinfo {author} {\bibfnamefont {P.}~\bibnamefont
  {Huovinen}}, \bibinfo {author} {\bibfnamefont {P.~F.}\ \bibnamefont {Kolb}},
  \bibinfo {author} {\bibfnamefont {U.~W.}\ \bibnamefont {Heinz}}, \bibinfo
  {author} {\bibfnamefont {P.~V.}\ \bibnamefont {Ruuskanen}}, \ and\ \bibinfo
  {author} {\bibfnamefont {S.~A.}\ \bibnamefont {Voloshin}},\ }\href@noop {}
  {\bibfield  {journal} {\bibinfo  {journal} {Phys. Lett.},\ }\textbf {\bibinfo
  {volume} {B503}},\ \bibinfo {pages} {58} (\bibinfo {year}
  {2001})}\BibitemShut {NoStop}%
\bibitem [{\citenamefont {Teaney}(2003)}]{Teaney:2003kp}%
  \BibitemOpen
  \bibfield  {author} {\bibinfo {author} {\bibfnamefont {D.}~\bibnamefont
  {Teaney}},\ }\Doi {10.1103/PhysRevC.68.034913} {\bibfield  {journal}
  {\bibinfo  {journal} {Phys. Rev.},\ }\textbf {\bibinfo {volume} {C68}},\
  \bibinfo {pages} {034913} (\bibinfo {year} {2003})}\BibitemShut {NoStop}%
\bibitem [{\citenamefont {Romatschke}\ and\ \citenamefont
  {Romatschke}(2007)}]{Romatschke:2007mq}%
  \BibitemOpen
  \bibfield  {author} {\bibinfo {author} {\bibfnamefont {P.}~\bibnamefont
  {Romatschke}}\ and\ \bibinfo {author} {\bibfnamefont {U.}~\bibnamefont
  {Romatschke}},\ }\Doi {10.1103/PhysRevLett.99.172301} {\bibfield  {journal}
  {\bibinfo  {journal} {Phys. Rev. Lett.},\ }\textbf {\bibinfo {volume} {99}},\
  \bibinfo {pages} {172301} (\bibinfo {year} {2007})}\BibitemShut {NoStop}%
\bibitem [{\citenamefont {Hama}\ \emph {et~al.}(2008)\citenamefont {Hama} \emph
  {et~al.}}]{Hama:2007dq}%
  \BibitemOpen
  \bibfield  {author} {\bibinfo {author} {\bibfnamefont {Y.}~\bibnamefont
  {Hama}} \emph {et~al.},\ }\href@noop {} {\bibfield  {journal} {\bibinfo
  {journal} {Phys. Atom. Nucl.},\ }\textbf {\bibinfo {volume} {71}},\ \bibinfo
  {pages} {1558} (\bibinfo {year} {2008})}\BibitemShut {NoStop}%
\bibitem [{\citenamefont {Song}\ and\ \citenamefont
  {Heinz}(2008)}]{Song:2007ux}%
  \BibitemOpen
  \bibfield  {author} {\bibinfo {author} {\bibfnamefont {H.}~\bibnamefont
  {Song}}\ and\ \bibinfo {author} {\bibfnamefont {U.~W.}\ \bibnamefont
  {Heinz}},\ }\href@noop {} {\bibfield  {journal} {\bibinfo  {journal} {Phys.
  Rev.},\ }\textbf {\bibinfo {volume} {C77}},\ \bibinfo {pages} {064901}
  (\bibinfo {year} {2008})}\BibitemShut {NoStop}%
\bibitem [{\citenamefont {Lacey}(2002)}]{Lacey:2001va}%
  \BibitemOpen
  \bibfield  {author} {\bibinfo {author} {\bibfnamefont {R.~A.}\ \bibnamefont
  {Lacey}},\ }\href@noop {} {\bibfield  {journal} {\bibinfo  {journal} {Nucl.
  Phys.},\ }\textbf {\bibinfo {volume} {A698}},\ \bibinfo {pages} {559}
  (\bibinfo {year} {2002})}\BibitemShut {NoStop}%
\bibitem [{\citenamefont {Snellings}(2002)}]{Snellings:2001nf}%
  \BibitemOpen
  \bibfield  {author} {\bibinfo {author} {\bibfnamefont {R.~J.~M.}\
  \bibnamefont {Snellings}},\ }\Doi {10.1016/S0375-9474(01)01364-1} {\bibfield
  {journal} {\bibinfo  {journal} {Nucl. Phys.},\ }\textbf {\bibinfo {volume}
  {A698}},\ \bibinfo {pages} {193} (\bibinfo {year} {2002})}\BibitemShut
  {NoStop}%
\bibitem [{\citenamefont {Adams}\ \emph {et~al.}(2004)\citenamefont {Adams}
  \emph {et~al.}}]{Adams:2003zg}%
  \BibitemOpen
  \bibfield  {author} {\bibinfo {author} {\bibfnamefont {J.}~\bibnamefont
  {Adams}} \emph {et~al.},\ }\Doi {10.1103/PhysRevLett.92.062301} {\bibfield
  {journal} {\bibinfo  {journal} {Phys. Rev. Lett.},\ }\textbf {\bibinfo
  {volume} {92}},\ \bibinfo {pages} {062301} (\bibinfo {year}
  {2004})}\BibitemShut {NoStop}%
\bibitem [{\citenamefont {Lacey}\ \emph {et~al.}(2009)\citenamefont {Lacey},
  \citenamefont {Taranenko},\ and\ \citenamefont {Wei}}]{Lacey:2009xx}%
  \BibitemOpen
  \bibfield  {author} {\bibinfo {author} {\bibfnamefont {R.~A.}\ \bibnamefont
  {Lacey}}, \bibinfo {author} {\bibfnamefont {A.}~\bibnamefont {Taranenko}}, \
  and\ \bibinfo {author} {\bibfnamefont {R.}~\bibnamefont {Wei}},\ }\href@noop
  {} { (\bibinfo {year} {2009})},\ \Eprint {http://arxiv.org/abs/0905.4368}
  {arXiv:0905.4368 [nucl-ex]} \BibitemShut {NoStop}%
\bibitem [{\citenamefont {Heinz}\ and\ \citenamefont
  {Wong}(2002)}]{Heinz:2002rs}%
  \BibitemOpen
  \bibfield  {author} {\bibinfo {author} {\bibfnamefont {U.~W.}\ \bibnamefont
  {Heinz}}\ and\ \bibinfo {author} {\bibfnamefont {S.~M.~H.}\ \bibnamefont
  {Wong}},\ }\Doi {10.1103/PhysRevC.66.014907} {\bibfield  {journal} {\bibinfo
  {journal} {Phys. Rev.},\ }\textbf {\bibinfo {volume} {C66}},\ \bibinfo
  {pages} {014907} (\bibinfo {year} {2002})}\BibitemShut {NoStop}%
\bibitem [{\citenamefont {Lacey}\ and\ \citenamefont
  {Taranenko}(2006)}]{Lacey:2006pn}%
  \BibitemOpen
  \bibfield  {author} {\bibinfo {author} {\bibfnamefont {R.~A.}\ \bibnamefont
  {Lacey}}\ and\ \bibinfo {author} {\bibfnamefont {A.}~\bibnamefont
  {Taranenko}},\ }\href@noop {} {\bibfield  {journal} {\bibinfo  {journal}
  {PoS},\ }\textbf {\bibinfo {volume} {CFRNC2006}},\ \bibinfo {pages} {021}
  (\bibinfo {year} {2006})}\BibitemShut {NoStop}%
\bibitem [{\citenamefont {Drescher}\ \emph {et~al.}(2007)\citenamefont
  {Drescher}, \citenamefont {Dumitru}, \citenamefont {Gombeaud},\ and\
  \citenamefont {Ollitrault}}]{Drescher:2007cd}%
  \BibitemOpen
  \bibfield  {author} {\bibinfo {author} {\bibfnamefont {H.-J.}\ \bibnamefont
  {Drescher}}, \bibinfo {author} {\bibfnamefont {A.}~\bibnamefont {Dumitru}},
  \bibinfo {author} {\bibfnamefont {C.}~\bibnamefont {Gombeaud}}, \ and\
  \bibinfo {author} {\bibfnamefont {J.-Y.}\ \bibnamefont {Ollitrault}},\
  }\href@noop {} {\bibfield  {journal} {\bibinfo  {journal} {Phys. Rev.},\
  }\textbf {\bibinfo {volume} {C76}},\ \bibinfo {pages} {024905} (\bibinfo
  {year} {2007})}\BibitemShut {NoStop}%
\bibitem [{\citenamefont {Xu}\ \emph {et~al.}(2008)\citenamefont {Xu},
  \citenamefont {Greiner},\ and\ \citenamefont {Stocker}}]{Xu:2007jv}%
  \BibitemOpen
  \bibfield  {author} {\bibinfo {author} {\bibfnamefont {Z.}~\bibnamefont
  {Xu}}, \bibinfo {author} {\bibfnamefont {C.}~\bibnamefont {Greiner}}, \ and\
  \bibinfo {author} {\bibfnamefont {H.}~\bibnamefont {Stocker}},\ }\Doi
  {10.1103/PhysRevLett.101.082302} {\bibfield  {journal} {\bibinfo  {journal}
  {Phys. Rev. Lett.},\ }\textbf {\bibinfo {volume} {101}},\ \bibinfo {pages}
  {082302} (\bibinfo {year} {2008})}\BibitemShut {NoStop}%
\bibitem [{\citenamefont {Greco}\ \emph {et~al.}(2008)\citenamefont {Greco},
  \citenamefont {Colonna}, \citenamefont {Di~Toro},\ and\ \citenamefont
  {Ferini}}]{Greco:2008fs}%
  \BibitemOpen
  \bibfield  {author} {\bibinfo {author} {\bibfnamefont {V.}~\bibnamefont
  {Greco}}, \bibinfo {author} {\bibfnamefont {M.}~\bibnamefont {Colonna}},
  \bibinfo {author} {\bibfnamefont {M.}~\bibnamefont {Di~Toro}}, \ and\
  \bibinfo {author} {\bibfnamefont {G.}~\bibnamefont {Ferini}},\ }\href@noop {}
  { (\bibinfo {year} {2008})},\ \Eprint {http://arxiv.org/abs/0811.3170}
  {arXiv:0811.3170 [hep-ph]} \BibitemShut {NoStop}%
\bibitem [{\citenamefont {Luzum}\ and\ \citenamefont
  {Romatschke}(2008)}]{Luzum:2008cw}%
  \BibitemOpen
  \bibfield  {author} {\bibinfo {author} {\bibfnamefont {M.}~\bibnamefont
  {Luzum}}\ and\ \bibinfo {author} {\bibfnamefont {P.}~\bibnamefont
  {Romatschke}},\ }\Doi {10.1103/PhysRevC.78.034915} {\bibfield  {journal}
  {\bibinfo  {journal} {Phys. Rev.},\ }\textbf {\bibinfo {volume} {C78}},\
  \bibinfo {pages} {034915} (\bibinfo {year} {2008})}\BibitemShut {NoStop}%
\bibitem [{\citenamefont {Chaudhuri}(2009)}]{Chaudhuri:2009hj}%
  \BibitemOpen
  \bibfield  {author} {\bibinfo {author} {\bibfnamefont {A.~K.}\ \bibnamefont
  {Chaudhuri}},\ }\href@noop {} { (\bibinfo {year} {2009})},\ \Eprint
  {http://arxiv.org/abs/0910.0979} {arXiv:0910.0979 [nucl-th]} \BibitemShut
  {NoStop}%
\bibitem [{\citenamefont {Song}\ and\ \citenamefont
  {Heinz}(2009)}]{Song:2008hj}%
  \BibitemOpen
  \bibfield  {author} {\bibinfo {author} {\bibfnamefont {H.}~\bibnamefont
  {Song}}\ and\ \bibinfo {author} {\bibfnamefont {U.~W.}\ \bibnamefont
  {Heinz}},\ }\Doi {10.1088/0954-3899/36/6/064033} {\bibfield  {journal}
  {\bibinfo  {journal} {J. Phys.},\ }\textbf {\bibinfo {volume} {G36}},\
  \bibinfo {pages} {064033} (\bibinfo {year} {2009})}\BibitemShut {NoStop}%
\bibitem [{\citenamefont {Hirano}\ \emph {et~al.}(2006)\citenamefont {Hirano},
  \citenamefont {Heinz}, \citenamefont {Kharzeev}, \citenamefont {Lacey},\ and\
  \citenamefont {Nara}}]{Hirano:2005xf}%
  \BibitemOpen
  \bibfield  {author} {\bibinfo {author} {\bibfnamefont {T.}~\bibnamefont
  {Hirano}}, \bibinfo {author} {\bibfnamefont {U.~W.}\ \bibnamefont {Heinz}},
  \bibinfo {author} {\bibfnamefont {D.}~\bibnamefont {Kharzeev}}, \bibinfo
  {author} {\bibfnamefont {R.}~\bibnamefont {Lacey}}, \ and\ \bibinfo {author}
  {\bibfnamefont {Y.}~\bibnamefont {Nara}},\ }\Doi
  {10.1016/j.physletb.2006.03.060} {\bibfield  {journal} {\bibinfo  {journal}
  {Phys. Lett.},\ }\textbf {\bibinfo {volume} {B636}},\ \bibinfo {pages} {299}
  (\bibinfo {year} {2006})}\BibitemShut {NoStop}%
\bibitem [{\citenamefont {Drescher}\ \emph {et~al.}(2006)\citenamefont
  {Drescher}, \citenamefont {Dumitru}, \citenamefont {Hayashigaki},\ and\
  \citenamefont {Nara}}]{Drescher:2006pi}%
  \BibitemOpen
  \bibfield  {author} {\bibinfo {author} {\bibfnamefont {H.-J.}\ \bibnamefont
  {Drescher}}, \bibinfo {author} {\bibfnamefont {A.}~\bibnamefont {Dumitru}},
  \bibinfo {author} {\bibfnamefont {A.}~\bibnamefont {Hayashigaki}}, \ and\
  \bibinfo {author} {\bibfnamefont {Y.}~\bibnamefont {Nara}},\ }\Doi
  {10.1103/PhysRevC.74.044905} {\bibfield  {journal} {\bibinfo  {journal}
  {Phys. Rev.},\ }\textbf {\bibinfo {volume} {C74}},\ \bibinfo {pages} {044905}
  (\bibinfo {year} {2006})}\BibitemShut {NoStop}%
\bibitem [{\citenamefont {Gombeaud}\ and\ \citenamefont
  {Ollitrault}(2009)}]{Gombeaud:2009ye}%
  \BibitemOpen
  \bibfield  {author} {\bibinfo {author} {\bibfnamefont {C.}~\bibnamefont
  {Gombeaud}}\ and\ \bibinfo {author} {\bibfnamefont {J.-Y.}\ \bibnamefont
  {Ollitrault}},\ }\href@noop {} { (\bibinfo {year} {2009})},\ \Eprint
  {http://arxiv.org/abs/0907.4664} {arXiv:0907.4664 [nucl-th]} \BibitemShut
  {NoStop}%
\bibitem [{\citenamefont {Alver}\ \emph {et~al.}(2007)\citenamefont {Alver}
  \emph {et~al.}}]{Alver:2006wh}%
  \BibitemOpen
  \bibfield  {author} {\bibinfo {author} {\bibfnamefont {B.}~\bibnamefont
  {Alver}} \emph {et~al.},\ }\Doi {10.1103/PhysRevLett.98.242302} {\bibfield
  {journal} {\bibinfo  {journal} {Phys. Rev. Lett.},\ }\textbf {\bibinfo
  {volume} {98}},\ \bibinfo {pages} {242302} (\bibinfo {year}
  {2007})}\BibitemShut {NoStop}%
\bibitem [{\citenamefont {Hirano}\ and\ \citenamefont
  {Nara}(2009)}]{Hirano:2009ah}%
  \BibitemOpen
  \bibfield  {author} {\bibinfo {author} {\bibfnamefont {T.}~\bibnamefont
  {Hirano}}\ and\ \bibinfo {author} {\bibfnamefont {Y.}~\bibnamefont {Nara}},\
  }\Doi {10.1103/PhysRevC.79.064904} {\bibfield  {journal} {\bibinfo  {journal}
  {Phys. Rev.},\ }\textbf {\bibinfo {volume} {C79}},\ \bibinfo {pages} {064904}
  (\bibinfo {year} {2009})}\BibitemShut {NoStop}%
\bibitem [{\citenamefont {Miller}\ \emph {et~al.}(2007)\citenamefont {Miller},
  \citenamefont {Reygers}, \citenamefont {Sanders},\ and\ \citenamefont
  {Steinberg}}]{Miller:2007ri}%
  \BibitemOpen
  \bibfield  {author} {\bibinfo {author} {\bibfnamefont {M.~L.}\ \bibnamefont
  {Miller}}, \bibinfo {author} {\bibfnamefont {K.}~\bibnamefont {Reygers}},
  \bibinfo {author} {\bibfnamefont {S.~J.}\ \bibnamefont {Sanders}}, \ and\
  \bibinfo {author} {\bibfnamefont {P.}~\bibnamefont {Steinberg}},\ }\Doi
  {10.1146/annurev.nucl.57.090506.123020} {\bibfield  {journal} {\bibinfo
  {journal} {Ann. Rev. Nucl. Part. Sci.},\ }\textbf {\bibinfo {volume} {57}},\
  \bibinfo {pages} {205} (\bibinfo {year} {2007})}\BibitemShut {NoStop}%
\bibitem [{\citenamefont {Lappi}\ and\ \citenamefont
  {Venugopalan}(2006)}]{Lappi:2006xc}%
  \BibitemOpen
  \bibfield  {author} {\bibinfo {author} {\bibfnamefont {T.}~\bibnamefont
  {Lappi}}\ and\ \bibinfo {author} {\bibfnamefont {R.}~\bibnamefont
  {Venugopalan}},\ }\Doi {10.1103/PhysRevC.74.054905} {\bibfield  {journal}
  {\bibinfo  {journal} {Phys. Rev.},\ }\textbf {\bibinfo {volume} {C74}},\
  \bibinfo {pages} {054905} (\bibinfo {year} {2006})}\BibitemShut {NoStop}%
\bibitem [{\citenamefont {Drescher}\ and\ \citenamefont
  {Nara}(2007)}]{Drescher:2007ax}%
  \BibitemOpen
  \bibfield  {author} {\bibinfo {author} {\bibfnamefont {H.-J.}\ \bibnamefont
  {Drescher}}\ and\ \bibinfo {author} {\bibfnamefont {Y.}~\bibnamefont
  {Nara}},\ }\Doi {10.1103/PhysRevC.76.041903} {\bibfield  {journal} {\bibinfo
  {journal} {Phys. Rev.},\ }\textbf {\bibinfo {volume} {C76}},\ \bibinfo
  {pages} {041903} (\bibinfo {year} {2007})}\BibitemShut {NoStop}%
\bibitem [{\citenamefont {Raman}\ and\ \citenamefont
  {Nestor}(1989)}]{Raman:1987yv}%
  \BibitemOpen
  \bibfield  {author} {\bibinfo {author} {\bibfnamefont {S.}~\bibnamefont
  {Raman}}\ and\ \bibinfo {author} {\bibfnamefont {C.~W.}\ \bibnamefont
  {Nestor}, \bibfnamefont {Jr.}},\ }\href@noop {} {\bibfield  {journal}
  {\bibinfo  {journal} {Atom. Data Nucl. Data Tabl.},\ }\textbf {\bibinfo
  {volume} {42}},\ \bibinfo {pages} {1} (\bibinfo {year} {1989})}\BibitemShut
  {NoStop}%
\bibitem [{\citenamefont {Moller}\ \emph {et~al.}(1995)\citenamefont {Moller},
  \citenamefont {Nix}, \citenamefont {Myers},\ and\ \citenamefont
  {Swiatecki}}]{Moller:1993ed}%
  \BibitemOpen
  \bibfield  {author} {\bibinfo {author} {\bibfnamefont {P.}~\bibnamefont
  {Moller}}, \bibinfo {author} {\bibfnamefont {J.~R.}\ \bibnamefont {Nix}},
  \bibinfo {author} {\bibfnamefont {W.~D.}\ \bibnamefont {Myers}}, \ and\
  \bibinfo {author} {\bibfnamefont {W.~J.}\ \bibnamefont {Swiatecki}},\
  }\href@noop {} {\bibfield  {journal} {\bibinfo  {journal} {Atom. Data Nucl.
  Data Tabl.},\ }\textbf {\bibinfo {volume} {59}},\ \bibinfo {pages} {185}
  (\bibinfo {year} {1995})}\BibitemShut {NoStop}%
\bibitem [{\citenamefont {Broniowski}\ \emph {et~al.}(2007)\citenamefont
  {Broniowski}, \citenamefont {Bozek},\ and\ \citenamefont
  {Rybczynski}}]{Broniowski:2007ft}%
  \BibitemOpen
  \bibfield  {author} {\bibinfo {author} {\bibfnamefont {W.}~\bibnamefont
  {Broniowski}}, \bibinfo {author} {\bibfnamefont {P.}~\bibnamefont {Bozek}}, \
  and\ \bibinfo {author} {\bibfnamefont {M.}~\bibnamefont {Rybczynski}},\
  }\href@noop {} {\bibfield  {journal} {\bibinfo  {journal} {Phys. Rev.},\
  }\textbf {\bibinfo {volume} {C76}},\ \bibinfo {pages} {054905} (\bibinfo
  {year} {2007})}\BibitemShut {NoStop}%
\bibitem [{\citenamefont {Back}\ \emph {et~al.}(2004)\citenamefont {Back} \emph
  {et~al.}}]{Back:2004dy}%
  \BibitemOpen
  \bibfield  {author} {\bibinfo {author} {\bibfnamefont {B.~B.}\ \bibnamefont
  {Back}} \emph {et~al.} (\bibinfo {collaboration} {PHOBOS}),\ }\Doi
  {10.1103/PhysRevC.70.021902} {\bibfield  {journal} {\bibinfo  {journal}
  {Phys. Rev.},\ }\textbf {\bibinfo {volume} {C70}},\ \bibinfo {pages} {021902}
  (\bibinfo {year} {2004})}\BibitemShut {NoStop}%
\bibitem [{\citenamefont {Shuryak}(2000)}]{Shuryak:1999by}%
  \BibitemOpen
  \bibfield  {author} {\bibinfo {author} {\bibfnamefont {E.~V.}\ \bibnamefont
  {Shuryak}},\ }\Doi {10.1103/PhysRevC.61.034905} {\bibfield  {journal}
  {\bibinfo  {journal} {Phys. Rev.},\ }\textbf {\bibinfo {volume} {C61}},\
  \bibinfo {pages} {034905} (\bibinfo {year} {2000})}\BibitemShut {NoStop}%
\bibitem [{\citenamefont {Li}(2000)}]{Li:1999bea}%
  \BibitemOpen
  \bibfield  {author} {\bibinfo {author} {\bibfnamefont {B.-A.}\ \bibnamefont
  {Li}},\ }\Doi {10.1103/PhysRevC.61.021903} {\bibfield  {journal} {\bibinfo
  {journal} {Phys. Rev.},\ }\textbf {\bibinfo {volume} {C61}},\ \bibinfo
  {pages} {021903} (\bibinfo {year} {2000})}\BibitemShut {NoStop}%
\bibitem [{\citenamefont {Heinz}\ and\ \citenamefont
  {Kuhlman}(2005)}]{Heinz:2004ir}%
  \BibitemOpen
  \bibfield  {author} {\bibinfo {author} {\bibfnamefont {U.~W.}\ \bibnamefont
  {Heinz}}\ and\ \bibinfo {author} {\bibfnamefont {A.}~\bibnamefont
  {Kuhlman}},\ }\Doi {10.1103/PhysRevLett.94.132301} {\bibfield  {journal}
  {\bibinfo  {journal} {Phys. Rev. Lett.},\ }\textbf {\bibinfo {volume} {94}},\
  \bibinfo {pages} {132301} (\bibinfo {year} {2005})}\BibitemShut {NoStop}%
\bibitem [{\citenamefont {Filip}\ \emph {et~al.}(2009)\citenamefont {Filip},
  \citenamefont {Lednicky}, \citenamefont {Masui},\ and\ \citenamefont
  {Xu}}]{Filip:2009zz}%
  \BibitemOpen
  \bibfield  {author} {\bibinfo {author} {\bibfnamefont {P.}~\bibnamefont
  {Filip}}, \bibinfo {author} {\bibfnamefont {R.}~\bibnamefont {Lednicky}},
  \bibinfo {author} {\bibfnamefont {H.}~\bibnamefont {Masui}}, \ and\ \bibinfo
  {author} {\bibfnamefont {N.}~\bibnamefont {Xu}},\ }\Doi
  {10.1103/PhysRevC.80.054903} {\bibfield  {journal} {\bibinfo  {journal}
  {Phys. Rev.},\ }\textbf {\bibinfo {volume} {C80}},\ \bibinfo {pages} {054903}
  (\bibinfo {year} {2009})}\BibitemShut {NoStop}%
\bibitem [{\ and\ \citenamefont {Adare}(2010)}]{Adare:2010ux}%
  \BibitemOpen
  \bibfield  {author} {\ and\ \bibinfo {author} {\bibfnamefont
  {A.}~\bibnamefont {Adare}} (\bibinfo {collaboration} {The PHENIX}),\
  }\href@noop {} { (\bibinfo {year} {2010})},\ \Eprint
  {http://arxiv.org/abs/1003.5586} {arXiv:1003.5586 [nucl-ex]} \BibitemShut
  {NoStop}%
\bibitem [{\citenamefont {Heinz}\ \emph {et~al.}(2009)\citenamefont {Heinz},
  \citenamefont {Moreland},\ and\ \citenamefont {Song}}]{Heinz:2009cv}%
  \BibitemOpen
  \bibfield  {author} {\bibinfo {author} {\bibfnamefont {U.~W.}\ \bibnamefont
  {Heinz}}, \bibinfo {author} {\bibfnamefont {J.~S.}\ \bibnamefont {Moreland}},
  \ and\ \bibinfo {author} {\bibfnamefont {H.}~\bibnamefont {Song}},\ }\Doi
  {10.1103/PhysRevC.80.061901} {\bibfield  {journal} {\bibinfo  {journal}
  {Phys. Rev.},\ }\textbf {\bibinfo {volume} {C80}},\ \bibinfo {pages} {061901}
  (\bibinfo {year} {2009})}\BibitemShut {NoStop}%
\bibitem [{\citenamefont {Kolb}\ \emph {et~al.}(2004)\citenamefont {Kolb},
  \citenamefont {Chen}, \citenamefont {Greco},\ and\ \citenamefont
  {Ko}}]{Kolb:2004gi}%
  \BibitemOpen
  \bibfield  {author} {\bibinfo {author} {\bibfnamefont {P.~F.}\ \bibnamefont
  {Kolb}}, \bibinfo {author} {\bibfnamefont {L.-W.}\ \bibnamefont {Chen}},
  \bibinfo {author} {\bibfnamefont {V.}~\bibnamefont {Greco}}, \ and\ \bibinfo
  {author} {\bibfnamefont {C.~M.}\ \bibnamefont {Ko}},\ }\Doi
  {10.1103/PhysRevC.69.051901} {\bibfield  {journal} {\bibinfo  {journal}
  {Phys. Rev.},\ }\textbf {\bibinfo {volume} {C69}},\ \bibinfo {pages} {051901}
  (\bibinfo {year} {2004})}\BibitemShut {NoStop}%
\end{thebibliography}%
\end{document}